\documentclass[%
reprint,
aps,
pra
]{revtex4-2}

\usepackage[usenames]{color}
\usepackage{colortbl}
\usepackage{graphicx}
\usepackage{dcolumn}
\usepackage{bm}
\usepackage{amsmath}
\usepackage{amssymb}
\usepackage[utf8]{inputenc}

\usepackage{comment}
\usepackage{dsfont}
\usepackage{xcolor}

\begin{document}

\preprint{APS/123-QED}

\title{Optical coprocessor based on spontaneous Brillouin scattering
}

\author{I. V. Vovchenko$^{1,3}$, A. A. Zyablovsky$^{1,2,3}$, A. A. Pukhov$^{1,2}$, E. S. Andrianov$^{1,2,3}$}
\affiliation{%
 $^1$Moscow Institute of Physics and Technology, 9 Institutskiy pereulok, Dolgoprudny 141700, Moscow region, Russia;
}%
\affiliation{
 $^2$Institute for Theoretical and Applied Electrodynamics, 13 Izhorskaya, Moscow 125412, Russia;
}
\affiliation{%
 $^3$Dukhov Research Institute of Automatics (VNIIA), 22 Sushchevskaya, Moscow 127055, Russia;
}%


\date{\today}

\begin{abstract}
Analog coprocessors for neural networks are an intensively developing field. 
They provide approximate results of computations for relatively low energy cost and at high speed.
We show that a set of ring resonators with Brillouin interaction between photons and phonons, being coupled to a waveguide, can be used to implement matrix-vector multiplication.
The input vector is formed by occupancies of the anti-Stokes optical modes pumped via spontaneous Brillouin scattering, i.e, scattering on thermal phonons. 
Brillouin scattering rates and coupling constants between ring resonators and the waveguide form the matrix.
The system allows for parallel computations in frequency band.
\end{abstract}

\maketitle


\section{Introduction}
Analog~\cite{shastri2021photonics,destras2023survey,ferreira2017progress,youngblood2022coherent,giamougiannis2023coherent,kovaios2025chip,pierangeli2021photonic,miscuglio2020photonic,hu2024diffractive,lin2018all,harris2018linear,bogaerts2020programmable,carolan2015universal,
aifer2024thermodynamic,melanson2025thermodynamic,abbas2021power,jia2019quantum,killoran2019continuous,kim2012functional,li2021hardware,9932877,xia2019memristive,ma2017darwin} and digital~\cite{wu2020energy,lazaro2007hardware,reuther2020survey} coprocessors are used in neural networks for acceleration of computations.
Digital coprocessors are more precise than analog coprocessors.
However, when only approximate results are needed, the usage of analog coprocessors is a preferable choice, as they are faster and show relatively low energy consumption~\cite{shastri2021photonics,destras2023survey,ferreira2017progress,li2021hardware,9932877,xia2019memristive,aifer2024thermodynamic,melanson2025thermodynamic,harris2018linear}.
The last is seen by some authors as a cornerstone of neural network computations in the future as data centers are extensively growing and their electricity consumption is evaluated to become of the order of electricity consumption of large countries in the following decade~\cite{ferreira2017progress,xia2019memristive,shehabi20242024,IEA}.

There are several main widely spread platforms for analog coprocessors: optical~\cite{shastri2021photonics,destras2023survey,ferreira2017progress,youngblood2022coherent,giamougiannis2023coherent,kovaios2025chip,pierangeli2021photonic,miscuglio2020photonic,hu2024diffractive,lin2018all,harris2018linear,bogaerts2020programmable,carolan2015universal}, thermodynamical~\cite{aifer2024thermodynamic,melanson2025thermodynamic}, quantum~\cite{abbas2021power,jia2019quantum,killoran2019continuous}, and memristor-based~\cite{kim2012functional,li2021hardware,9932877,xia2019memristive,ma2017darwin}.
Optical coprocessors implement linear operations in optical crossbar structures~\cite{shastri2021photonics,destras2023survey,youngblood2022coherent,giamougiannis2023coherent,kovaios2025chip}, meshes of Mach–Zehnder interferometers~\cite{harris2018linear,bogaerts2020programmable,carolan2015universal}, and by means of free-space optics~\cite{pierangeli2021photonic,hu2024diffractive,lin2018all} using bosonic modes.
The frequency band in this case can be used as another degree of freedom, providing operations with tensors of higher dimension~\cite{miscuglio2020photonic,youngblood2022coherent,destras2023survey}.
Thermodynamical coprocessors use natural relaxation of a system to its equilibrium states to conduct computations~\cite{aifer2024thermodynamic,melanson2025thermodynamic}.
The answer is stored in parameters of the system stationary state~\cite{aifer2024thermodynamic,melanson2025thermodynamic,lipka2024thermodynamic}.
Quantum coprocessors use entanglement to increase the number of computations per second~\cite{abbas2021power,jia2019quantum,killoran2019continuous}.
Memristor coprocessors use crossbar circuits to implement linear algebra operations~\cite{kim2012functional,li2021hardware,9932877,xia2019memristive,ma2017darwin,chen2023full}.


In addition, full-fledged development of analog neural networks requires usage of non-linear elements to realize activation functions~\cite{destras2023survey,slinkov2025all,chen2023full}.
In optical neural networks, such elements usually employ electro-optical effects, two-photon absorption, Kerr non-linearity, free carrier absorption, and dispersion impact~\cite{destras2023survey}.
Besides, non-linear photon-phonon interaction
that induces the Brillouin scattering~\cite{sipe2016hamiltonian,lisyansky2024quantum,shishkov2021cascade,damzen2003stimulated,rodrigues2023stimulated} in substances like SiO$_2$~\cite{heiman1979brillouin}, silica~\cite{bahl2012observation}, LiNbO$_3$~\cite{rodrigues2023stimulated}, and Si$_3$N$_4$~\cite{chauhan2020evidence} shows great potential in achieving different practically feasible all-optical activation functions that also can be switched in an all-optical manner at a single platform~\cite{slinkov2025all}.
Thus, it would be convenient to use a similar mechanism to implement linear operations such as matrix-vector multiplication.

In this work, we show that a set of ring resonators with Brillouin interaction between photons and phonons, coupled to a waveguide, can be used as an analog coprocessor implementing linear operations.
Anti-Stokes optical modes in the ring resonators are pumped via spontaneous Brillouin scattering on thermal phonons.
Coupling of the ring resonators' optical modes to the waveguide excites its corresponding resonant modes.
We show that the stationary intensity of the waveguide's mode resonant to the anti-Stokes mode of the ring resonator is proportional to the scalar product of a vector composed of the phonon modes' occupancies and a vector formed by the spontaneous Brillouin scattering rates and the coupling constants between the ring resonators and the waveguide.
Hence, being copied, this system can implement multiplication of a vector by a matrix.
We show that the considering system allows for parallel computations in frequency band.
The computational time weakly depends on the input vector dimension and is of the order of the time needed for the system to reach its stationary state.

\section{The model}
\begin{figure}
    \centering
    \includegraphics[width=0.96\linewidth]{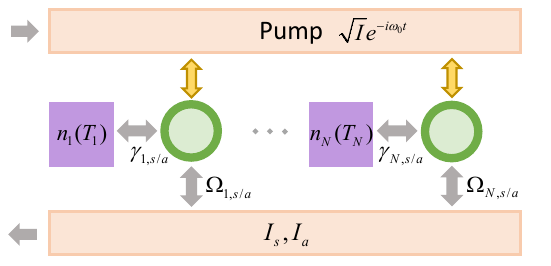}
    \caption{Schematic representation of the system under consideration
    ($I_a=\langle\hat{a}_a^\dag \hat{a}_a\rangle$, $I_s=\langle\hat{a}_s^\dag \hat{a}_s\rangle$).}
    \label{Phonons}
\end{figure}

Let consider a system of $N$ ring resonators with a resonant frequency $\omega_0$ and Brillouin interaction between photons and phonons.
The ring resonators are coupled to an optical waveguide, see Fig.~\ref{Phonons}, i.e., optical modes of the ring resonators interact with the resonators' internal reservoirs of thermal phonons.
These optical modes are coupled to optical modes of the waveguide.
A pumping at frequency $\omega_0$ with dimensionless intensity $I$ (measured in units of $\omega_0$) is introduced in the waveguide.
We consider that the dynamics of modes with $\omega_0$ frequency are completely determined by this pumping. 
Therefore, we exclude these degrees of freedom from further consideration.
The Brillouin scattering results in excitation of Stokes and anti-Stokes modes in ring resonators with corresponding frequencies $\omega_s$ and $\omega_a$ (the $\omega_0-\omega_s\approx\omega_a-\omega_0$, see Appendix).
Let the frequency step $\Delta\omega$ in ring resonators be matching the Stokes shift, i.e., $\Delta\omega=\omega_a-\omega_0$.
The Hamiltonian of the system reads $\hat{H}=\hat{H}_S +\hat{H}_R +\hat{H}_{SR}$, where
\begin{gather}
\nonumber
        \hat{H}_S
        =
        \omega_s\hat{a}_s^\dag \hat{a}_s
        +\omega_a\hat{a}_a^\dag \hat{a}_a
        +\sum_{k=1}^N 
            \left(
                \omega_s \hat{b}^\dag_{k,s} \hat{b}_{k,s}
                +\omega_a \hat{b}^\dag_{k,a} \hat{b}_{k,a}
            \right)
        \\ \nonumber
        + \sum_{k=1}^N 
            \Omega_{k,s}(\hat{a}_s^\dag\hat{b}_{k,s} +\hat{b}_{k,s}^\dag \hat{a}_s)
        +\sum_{k=1}^N 
            \Omega_{k,a}(\hat{a}_a^\dag\hat{b}_{k,a} +\hat{b}_{k,a}^\dag \hat{a}_a),
        \\ \label{H_fre_L0}
        \hat{H}_R=\sum_{k=1}^N \sum_{q_k} \omega_{k,q_k} \hat c_{k,q_k}^\dag \hat c_{k,q_k}, 
        \\ \nonumber
        \hat{H}_{SR}
        =
        \sqrt{I}\sum_{k=1}^N \sum_{q_k} \chi_{k,q_k}(\hat{b}_{k,a}^\dag \hat c_{k,q_k} e^{-i\omega_0 t} + \hat c_{k,q_k}^\dag \hat{b}_{k,a} e^{i\omega_0 t})
        \\ \nonumber
        +\sqrt{I}\sum_{k=1}^N \sum_{q_k} \theta_{k,q_k}(\hat{b}_{k,s}^\dag \hat c_{k,q_k}^\dag e^{-i\omega_0 t} + \hat c_{k,q_k} \hat{b}_{k,s}  e^{i\omega_0 t}).
\end{gather}
Here indices $s$ and $a$ stand for Stokes and anti-Stokes modes,
$\hat{b}_{k,s/a}$ is the annihilation operator of the corresponding optical mode in the $k$-th ring resonator, $\hat{a}_{s/a}$ is the annihilation operator of the waveguide mode resonant to the corresponding mode of ring resonators (further, waveguide's Stokes and anti-Stokes modes), $\Omega_{k,s/a}$ is the coupling constant between the corresponding waveguide's mode and the corresponding optical mode of the $k$-th ring resonator, $\hat{c}_{k,q_k}$ is the annihilation operator of the $q_k$-th thermal phonon mode with frequency $\omega_{k,q_k}$ in the $k$-th ring resonator's phonon reservoir, $\chi_{k,q_k}$ is the coupling constant between the anti-Stokes optical mode of the $k$-th ring resonator and $q_k$-th thermal phonon mode in the $k$-th ring resonator's phonon reservoir, $\theta_{k,q_k}$ is the coupling constant between the Stokes optical mode of the $k$-th ring resonator and $q_k$-th thermal phonon mode in the $k$-th ring resonator's phonon reservoir.

The interaction with phonons' reservoirs can be described via the Lindblad master equation on reduced density matrix $\hat{\rho}_S(t)$~\cite{carmichael2009open,breuer2002theory,shishkov2019relaxation}.
It governs the dynamics of an open quantum system (OQS), consisting of the mentioned optical modes (see Appendix).
\begin{gather}
\nonumber
    \frac{\partial \hat \rho_S}{\partial t}
    =
    i\left[
        \hat \rho, \hat{H}_S
    \right]
    + I\sum_{k=1}^N D_{k,a}[\hat\rho_S]
    + I\sum_{k=1}^N D_{k,s}[\hat\rho_S]
    \\ \label{master_eq}
    + D_{a}[\hat\rho_S]
    + D_{s}[\hat\rho_S].
\end{gather}

Here
\begin{align}
    D_{k,a}[\hat\rho_S]
    =&
    0.5\gamma_k(\omega_\phi)
    (n_k(\omega_\phi,T_k)+1)\hat L[\hat b_{k,a},\hat b^\dag_{k,a}] 
    \\ \nonumber
    +&0.5\gamma_k(\omega_\phi)
    n_k(\omega_\phi,T_k)\hat L[\hat b^\dag_{k,a},\hat b_{k,a}],
    \\ \nonumber
    D_{k,s}[\hat\rho_S]
    =&
    0.5\gamma_k(\omega_\phi)
    (n_k(\omega_\phi,T_k)+1)\hat L[\hat b_{k,s}^\dag,\hat b_{k,s}] 
    \\ \nonumber
    +&0.5\gamma_k(\omega_\phi)
    n_k(\omega_\phi,T_k)\hat L[\hat b_{k,s},\hat b^\dag_{k,s}],
    \\ \nonumber
    D_{a}[\hat\rho_S]
    =&
    0.5\gamma_a(\omega_a) (n_a(\omega_a,T_a)+1)\hat L[\hat a_a,\hat a_a^\dag] 
    \\ \nonumber
    +&0.5\gamma_a(\omega_a) n_a(\omega_a,T_a)\hat L[\hat a_a^\dag,\hat a_a],
    \\ \nonumber
    D_{s}[\hat\rho_S]
    =&
    0.5\gamma_s(\omega_s) (n_a(\omega_s,T_a)+1)\hat L[\hat a_s,\hat a_s^\dag] 
    \\ \nonumber
    +&0.5\gamma_s(\omega_s) n_a(\omega_s,T_a)\hat L[\hat a_s^\dag,\hat a_s].
\end{align}
Here $\omega_\phi=\omega_a-\omega_0\approx-\omega_s+\omega_0$ is the frequency of the phonons involved in the Brillouin scattering ($\omega_\phi$ coincides with $\Delta\omega$), superoperator $\hat L [\hat X, \hat Y] = 2\hat X \hat \rho_S \hat Y-\hat Y \hat X \hat\rho_S -\hat\rho_S \hat Y \hat X$, $n_{k}(\tilde\omega,T_{k})=(e^{\tilde\omega/T_{k}}-1)^{-1}$ is the occupancy of the phonon reservoir in $k$-th ring resonator at frequency $\tilde\omega$ ($T_{k}$ is the temperature of this reservoir), $\gamma_{k}(\tilde\omega) = \pi g_k\left(\tilde\omega\right) |\chi_{k,\mathbf{q}}|^2$ we further address as Brillouin scattering rate in $k$-th ring resonator, where $g_j\left(\tilde\omega\right)$ is the density of states in $k$-th ring resonator's phonon reservoir, $\chi_{k,\mathbf{q}}$ is the coupling constant $\chi_{k,q_k}$ at frequency $\omega_{q_k}=\tilde\omega$.
In this master equation we also consider the dissipation of the waveguide modes, i.e., terms $D_{a}[\hat{\rho}_S]$ and $D_{s}[\hat{\rho}_S]$.
The $n_{a}(\tilde\omega,T_{a})=(e^{\tilde\omega/T_{a}}-1)^{-1}$ is the occupancy at frequency $\tilde{\omega}$ of the dissipative reservoir that interacts with the waveguide's modes ($T_{a}$ is the temperature of this reservoir),
$\gamma_s(\tilde{\omega})$ and $\gamma_a(\tilde{\omega})$ are dissipation rates of the waveguide modes resonant to the Stokes and anti-Stokes modes of the ring resonators, respectively.

Using the equations $d\langle \hat A\rangle/dt=\rm{tr({ \dot{\hat {\rho}}_S } \hat A)}$ and
${\rm tr}\left(L[{\hat X},\hat Y]\hat A\right)=\langle {\hat Y[ {\hat A,\hat X} ]} \rangle  + \langle {[ {\hat Y,\hat A} ]\hat X} \rangle$~\cite{vovchenko2021model,vovcenko2023energy,vovchenko2024transient}
we get the system of equations describing dynamics of occupancies in anti-Stokes modes ($I_a=\langle\hat{a}_a^\dag \hat{a}_a\rangle$)
\begin{align}
\nonumber
    \frac{\partial I_a}{\partial t}
    =
    -&i\sum_{k=1}^N \Omega_{k,a}\left(\langle\hat{a}^\dag_a \hat{b}_{k,a}\rangle-\langle\hat{b}_{k,a}^\dag \hat{a}_a\rangle\right) 
    - \gamma_a(\omega_a) I_a, 
    \\ \label{Diff_occ}
    \frac{\partial \langle\hat{b}_{k,a}^\dag\hat{b}_{k,a}\rangle}{\partial t}
    &= 
    i \Omega_{k,a}\left(\langle\hat{a}^\dag_a \hat{b}_{k,a}\rangle-\langle\hat{b}_{k,a}^\dag \hat{a}_a\rangle\right) 
    \\ \nonumber
    &- I\gamma_k(\omega_\phi) \langle \hat{b}_{k,a}^\dag \hat{b}_{k,a} \rangle 
    + I\gamma_k(\omega_\phi) n_k(\omega_\phi,T_k),
    \\ \nonumber
    \frac{\partial \langle\hat{a}_a^\dag\hat{b}_{k,a}\rangle}{\partial t}
    &= 
    i \Omega_{k,a} \left(\langle\hat{b}_{k,a}^\dag\hat{b}_{k,a}\rangle- \langle\hat{a}_a^\dag\hat{a}_a\rangle\rangle\right) 
    \\ \nonumber
    +i\sum_{j\ne k}^N &\Omega_{j,a}\langle\hat{b}_{j,a}^\dag\hat{b}_{k,a}\rangle
    -\frac{\gamma_a(\omega_a)+I\gamma_k(\omega_\phi)}{2} \langle\hat{a}_a^\dag\hat{b}_{k,a}\rangle, 
    \\ \nonumber
    \frac{\partial \langle\hat{b}_{j,a}^\dag\hat{b}_{k,a}\rangle}{\partial t}
    &=
    i\left(\Omega_{j,a}\langle\hat{a}_a^\dag\hat{b}_k\rangle-\Omega_{k,a}\langle\hat{b}_{j,a}^\dag\hat{a}_a\rangle\right)
    \\ \nonumber
    &-I\frac{\gamma_j(\omega_\phi)+\gamma_k(\omega_\phi)}{2}\langle\hat{b}_{j,a}^\dag\hat{b}_{k,a}\rangle.
\end{align}
Here we consider $n_a(\omega_a)=0$ as $\omega_a/T_a \gg 1$.
Note that cross-terms $\langle\hat{b}_{j,a}^\dag\hat{b}_{k,a}\rangle$ describing energy exchange between modes that do not interact directly affect the dynamics of the OQS.
However, if $\Omega_{k,a}\ll I\gamma_k(\omega_\phi)$, these cross-terms are negligible while condition $\sum_{k=1}^N\Omega_{k,a}^2 /I^2\gamma_k^2(\omega_\phi)\ll 1$ is satisfied (see Appendix).
If these conditions are fulfilled and $\gamma_a(\omega_a)\ll I\gamma_k(\omega_\phi)$, one gets $\overline{\langle\hat{b}_{k,a}^\dag \hat b_{k,a}\rangle}\approx n_k(\omega_\phi)$ and



\begin{figure}
    \centering
    \includegraphics[width=0.88\linewidth]{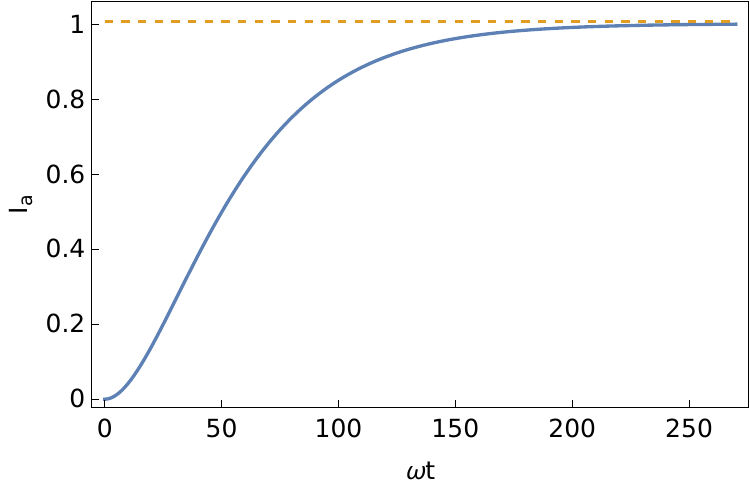}
    \caption{Dynamics of the waveguide's anti-Stokes mode occupancy $I_a$ (solid line) and $\tilde n (\omega_\phi,T_{1,\cdots,N})$ (dashed line) for $N=50$, $\Omega_{k,a}/\omega_0$ being set randomly between $5\cdot 10^{-4}$ and $5\cdot 10^{-3}$, $\gamma_\phi=0.1\omega_0$, $\gamma_a=5\cdot 10^{-4}\omega_0$, $n_k(\omega_\phi)$ being set randomly between $0.1$ and $2.0$, $(I_a/\omega_0)|_{t=0}=0$, $\omega=\omega_0$.}
    \label{aa_st0}
\end{figure}

\begin{equation}\label{stat_aa}
     \overline{I}_a
     \approx
     \frac{\sum\limits_{k=1}^N r_k (\omega_\phi,\omega_a,I,\Omega_{k,a}) n_k(\omega_\phi,T_k)}
        {\sum\limits_{k=1}^N r_k(\omega_\phi,\omega_a,I,\Omega_{k,a}) +\gamma_a (\omega)}.
\end{equation}
Here,  $\overline{\,\cdot\,}$ denotes the stationary value of the variable, and $r_k(\omega_\phi,\omega_a,I,\Omega_{k,a})=4\Omega_{k,a}^2/ (I\gamma_k(\omega_\phi)+\gamma_a (\omega_a))$.
Considering that all $\gamma_k(\omega_\phi)$ are the same, and varying the intensity $I$ of the pumping wave, we can achieve $\sum_{k=1}^N \Omega_{k}^2 \gg I\gamma_k(\omega_\phi) \gamma_a(\omega_a)$. 
Then, Eq.~(\ref{stat_aa}) transforms to
\begin{equation}
     \overline{I}_a
     \approx
     \sum\limits_{k=1}^N \Omega_{k,a}^2 n_k(\omega_\phi,T_k)
     /\sum\limits_{k=1}^N \Omega_{k,a}^2
     \equiv
     \tilde n (\omega_\phi,T_{1,\cdots,N}).
\end{equation}

\begin{figure}
    \centering
    \includegraphics[width=0.88\linewidth]{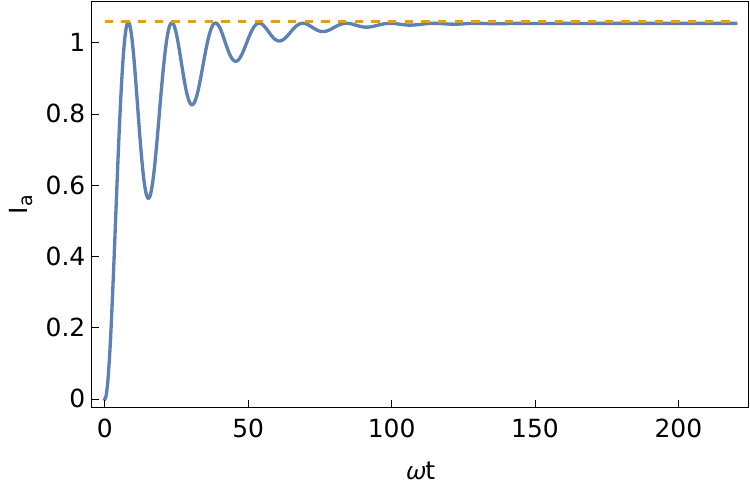}
    \caption{Dynamics of the waveguide's anti-Stokes mode occupancy $I_a$ (solid line) and $\tilde n (\omega_\phi,T_{1,\cdots,N})$ (dashed line) for $N=50$, $\Omega_{k,a}/\omega_0$ being set randomly between $5\cdot 10^{-3}$ and $5\cdot 10^{-2}$, $\gamma_\phi=0.1\omega_0$, $\gamma_a=5\cdot 10^{-4}\omega_0$, $n_k(\omega_\phi)$ being set randomly between $0.1$ and $2.0$, $(I_a/\omega_0)|_{t=0}=0$, $\omega=\omega_0$.}
    \label{aa_st}
\end{figure}

It is seen that $\overline{I}_a$ is equal to the weighted occupancy at frequency $\omega_\phi$ among phonon reservoirs, i.e., $\tilde n (\omega_\phi,T_{1,\cdots,N})$.
Weight coefficients are equal to weighted squares of coupling coefficients between the anti-Stokes modes of the ring resonators and the anti-Stokes mode of the waveguide.
The dynamics of $I_a$ are presented in Fig.~\ref{aa_st0} and Fig.~\ref{aa_st}.
It is seen that $I_a$ indeed tends to the value $\tilde n (\omega_\phi,T_{1,\cdots,N})$ while $t\rightarrow +\infty$.
Notably, $I_a$ exhibits non-oscillating (Fig.~\ref{aa_st0}) and oscillating (Fig.~\ref{aa_st}) dynamics depending on values of coupling constants $\Omega_k$.
Such type of behavior is typical for OQSs with exceptional points~\cite{vovcenko2023energy,sergeev2022environment,sergeev2025spontaneous}.

If $\gamma_a(\omega)\gg I\gamma_k(\omega_\phi)$, we have
\begin{equation}\label{aa_g}
     \overline{I}_a
     \approx
     \frac{4}{\gamma_a^2(\omega_a)}\sum\limits_{k=1}^N \Omega_{k,a}^2 n_k(\omega_\phi,T_k).
\end{equation}

\begin{figure}
    \centering
    \includegraphics[width=0.88\linewidth]{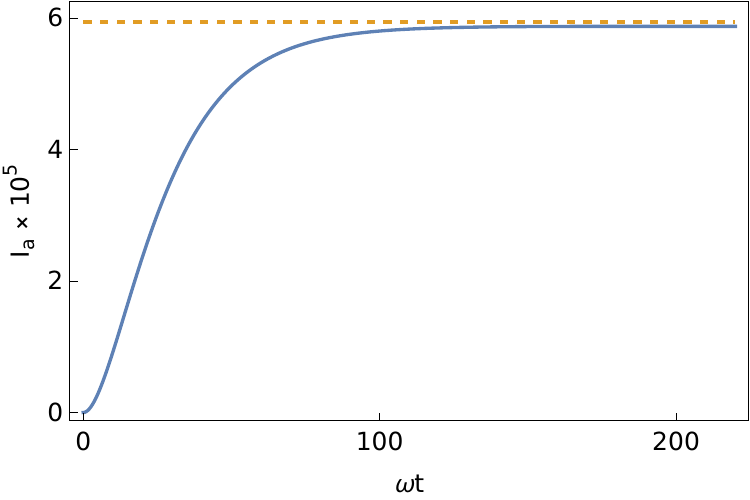}
    \caption{Dynamics of the waveguide's anti-Stokes mode occupancy $I_a$ (solid line) and value $4 \sum_k \Omega_{k,a}^2 n_k (\omega_\phi) / \gamma_a^2$ (dashed line) for $N=50$, $\Omega_{k,a}/\omega_0$ being set randomly between $10^{-5}$ and $10^{-4}$, $\gamma_\phi=10^{-3}\omega_0$, $\gamma_a=10^{-1}\omega_0$, $n_k(\omega_\phi)$ being set randomly between $0$ and $2$, $(I_a/\omega_0)|_{t=0}=0$, $\omega=\omega_0$.}
    \label{aa_st_g}
\end{figure}

In this case $\overline{I}_a$ is proportional to the scalar product of the vector composed of $\Omega_{k,a}^2$ and phonon reservoirs' occupancies $n_k(\omega_\phi,T_k)$ at frequency $\omega_\phi$.
The dynamics of $I_a$ is presented in Fig.~\ref{aa_st_g}.
It is seen that $I_a$ indeed tends to the value on the right-hand side of Eq.~(\ref{aa_g}) while $t\rightarrow +\infty$.

\section{Implementation of matrix-vector multiplication}
Let consider an array consisting of $M$ copies of the above system, but with different coupling constants between the waveguides and the ring resonators,
i.e., $\Omega_{k,a}^{(m)}$, where $m$ denotes the belonging of this coupling constant to the $m$-th element of the array.
Using this array, one can implement multiplication of an $M\times N$ matrix and an $N$-dimensional vector.
Indeed,
\begin{gather}
\nonumber
    \left[
    \begin{array}{cc}
        \overline{I}_{a,1}\\
        \vdots\\
        \overline{I}_{a,M}
    \end{array}
    \right]
    \approx
    \left[
    \begin{array}{ccc}
        r_{11} & \cdots  & r_{1N}\\
        \vdots & \ddots  & \vdots \\
        r_{M,1} & \cdots & r_{M,N}
    \end{array}
    \right]
    \left[
    \begin{array}{cc}
        n_1(\omega_\phi,T_1)\\
        \vdots\\
        n_N(\omega_\phi,T_N)
    \end{array}
    \right],
    \\
    r_{mk}
    =
    \frac{r_k\left(\omega_\phi,\omega_a,I,\Omega_{k,a}^{(m)}\right)}{\sum\limits_{k=1}^N r_k\left(\omega_\phi,\omega_a,I,\Omega_{k,a}^{(m)}\right)}.
\end{gather}
Here $\overline{I}_{a,m}$ is the stationary occupancy of the anti-Stokes mode in the waveguide of the $m$-th element of the array.
It is seen that the occupancies of phonon reservoirs form an $N$-dimensional input vector, values $r_{mk}$ form the matrix that is multiplied by the input vector, and the result of this operation is stored in the vector formed by stationary occupancy of the anti-Stokes modes in the waveguides.

This procedure of matrix-vector multiplication takes time that is needed for the OQS to reach its stationary state.
This time depends on $\gamma_a(\omega_a)$, $I\gamma_\phi(\omega_\phi)$, and it weakly depends on the number of OQS reservoirs', unless dark states are not involved in the dynamics~\cite{vovcenko2021dephasing,willingham2011energy,jenkins2017many,zhou2011tunable,meng2023strong}.
Usually, the OQS reaches its stationary state in nanoseconds~\cite{vovcenko2021dephasing,vovchenko2025thermodynamic,carmichael2009open}.

We can remove the dependence of $r_{mk}$ on $\gamma_k(\omega_\phi)$, if $\gamma_a(\omega_a)\ll I\gamma_k(\omega_\phi)$ and $\sum_{k=1}^N \Omega_{k,a}^2 \gg I\gamma_k(\omega_\phi) \gamma_a(\omega_a)$, as it is shown above.
Then, we get
\begin{gather}
\nonumber
    \left[
    \begin{array}{cc}
        \overline{I}_{a,1}\\
        \vdots\\
        \overline{I}_{a,M}
    \end{array}
    \right]
    \approx
    \left[
    \begin{array}{ccc}
        p_{11} & \cdots  & p_{1N}\\
        \vdots & \ddots  & \vdots \\
        p_{M,1} & \cdots & p_{M,N}
    \end{array}
    \right]
    \left[
    \begin{array}{cc}
        n_1(\omega_\phi,T_1)\\
        \vdots\\
        n_N(\omega_\phi,T_N)
    \end{array}
    \right],
    \\ 
    \text{where }
    p_{mk}
    =
    \left(\Omega_{k,a}^{(m)}\right)^2/{\sum\limits_{q=1}^N\left(\Omega_{q,a}^{(m)}\right)^2}.
\end{gather}

This matrix is a stochastic one, i.e., it is a transient matrix of a Markov process~\cite{norris1998markov,douc2018markov,freedman2012markov,agbinya2022applied,tweedie2001markov}, as $\sum_{k=1}^N p_{mk}=1$.
Hence, in this regime, the considering system can implement multiplication of a stochastic matrix by a positive vector.
Such an operation is frequently used in neural networks~\cite{agbinya2022applied,sigaud2013markov,bishop2006pattern,mustafa2019comparative,sutton1997significance}.

In the opposite case, when $I\gamma_k(\omega_\phi)\ll \gamma_a(\omega_a)$, we get
\begin{gather}
\nonumber
    \left[
    \begin{array}{cc}
        \overline{I}_{a,1}\\
        \vdots\\
        \overline{I}_{a,M}
    \end{array}
    \right]
    \approx
    \left[
    \begin{array}{ccc}
        \tilde{p}_{11} & \cdots  & \tilde{p}_{1N}\\
        \vdots & \ddots  & \vdots \\
        \tilde{p}_{M,1} & \cdots & \tilde{p}_{M,N}
    \end{array}
    \right]
    \left[
    \begin{array}{cc}
        n_1(\omega_\phi,T_1)\\
        \vdots\\
        n_N(\omega_\phi,T_N)
    \end{array}
    \right],
    \\
    \text{where }
    \tilde{p}_{mk}
    =
    4\left(\Omega_{k,a}^{(m)}\right)^2/\gamma_a^2(\omega_a).
\end{gather}
In this case the matrix is an arbitrary positive matrix with $0\le\tilde{p}_{mk}\ll 1$.
Amplifying the elements of the output vector, one can get multiplication with an arbitrary positive matrix instead.

In all regimes considered above, we can extend the class of matrices involved in computations, i.e., matrices with negative elements can be used.
For that we need two arrays described above.
One array will process the matrix composed of the initial matrix's positive elements and zeros in place of negative elements.
The second array will process the matrix composed of the initial matrix's negative elements and zeros in place of positive elements.
The second result should be subtracted from the first one to get the answer.

\section{Parallel computations in frequency band}
As it was mentioned earlier, the frequency band can be used as another degree of freedom in optical computations.
This can sufficiently increase the number of computations per second.
In this section we discuss the possibility of using this feature in the case of the considering system.

The frequency $\omega_0$ is a resonant frequency of ring resonators.
Let the frequency step in ring resonators be fixed $\Delta\omega=\omega_a-\omega_0\approx2\omega_0 v/c$, where $v$ and $c$ are the speed of sound and the speed of light in the material of the ring resonator, respectively.
It is seen that $\omega_a-\omega_0\sim\omega_0$,
hence, it is impossible to design a ring resonator that will precisely match the anti-Stokes modes at all its frequencies.

Due to dissipation, the lines of resonators have finite width.
Let denote its characteristic value as $\delta\omega$.
To implement the procedure described above at several frequencies simultaneously, the anti-Stokes modes should fall inside the linewidth of resonators.
Let estimate the number of modes that can be involved in this process.
For that, let consider the anti-Stokes mode formed from scattering of the ring resonator's mode with frequency $\omega_{0,K}=\omega_0+K\Delta\omega$, where $K$ is an integer.
The frequency of the anti-Stokes mode formed by this scattering equals $\omega_{a,K}\approx\omega_{0,K}+\omega_{0,K}2v/c$ (see Appendix).
Hence,
\begin{gather}
    \omega_{a,K}-(\omega_0+K\Delta\omega)\approx(\omega_0+K\Delta\omega)2v/c\approx
    \\ \nonumber
    \approx
    \omega_0 2v/c+K\omega_0(2v/c)^2.
\end{gather}

It is seen that the term that prevents frequency matching between anti-Stokes modes and the ring resonators modes is of the order $v^2/c^2$.
Then for the largest possible $K_{max}$ holds $K_{max}\omega_0 v^2/c^2\sim\delta\omega$.
Hence, $K_{max}\sim1/Q(v/c)^2$, where $Q\sim\omega_0/\delta\omega$ is the Q-factor of ring resonators.

To implement parallel optical computations in frequency band, the width of the involved spectral lines should be much less than the distance between them in the frequency band.
Thus, the value of $Q$ is restricted by the condition $\delta\omega\ll\Delta\omega=\omega_0 2v/c$.
Let $v/c\sim 10^{-5}$.
Hence, we need $Q\gg10^5$.
For $Q=10^6\div 10^8$, we get $K_{max}\sim 10^2\div 10^4$ ($K_{max}$ equals the maximal possible number of parallel computational channels).

\section{Conclusion and discussion}
In this work we have shown that a system of ring resonators with Brillouin interaction between photons and phonons, being coupled to a waveguide, can be used as an analog coprocessor.
The anti-Stokes optical modes in ring resonators are pumped via spontaneous Brillouin scattering on thermal phonons.
Coupling of the ring resonators optical modes to the waveguide modes excites the waveguide's modes resonant to the Stokes and anti-Stokes modes of the ring resonators.
We show that the stationary intensity of the waveguide's mode resonant to the anti-Stokes mode of the ring resonators is proportional to the scalar product of a vector composed of the phonon modes' occupancies and a vector formed by the spontaneous Brillouin scattering rates and the coupling constants between ring resonators and the waveguide.
Hence, being copied, this system can implement multiplication of a vector by a matrix.

The computational time weakly depends on the input vector dimension and is of the order of the time needed for the system to reach its stationary state.
This time is estimated to be at the level of nanoseconds.

Also, we show that the considering system allows for parallel computations at many modes of ring resonators simultaneously.
The number of modes that can be used in parallel computations simultaneously is restricted by the frequency dependence of the anti-Stokes modes' frequency shifts.
The rate of computations can be increased by $10^2\div 10^4$ times using this approach.

In the end we mention that usage of the non-linear process of Brillouin scattering for implementation of matrix-vector multiplication is a promising approach to computations in optical neural networks.
The Brillouin scattering process is already adapted for the realization of different all-optical activation functions that can be switched in an all-optical manner~\cite{slinkov2025all}.
Implementation of both linear and non-linear operations that are performed in neural networks at a platform sharing the same physical principle can potentially reduce the energy cost of computations and increase their speed.

\textbf{Acknowledgment.}
A.A.Z. and E.S.A. acknowledge the support of the Foundation for the Advancement of Theoretical Physics and Mathematics BASIS.


\bibliography{CwCM}

\section*{Appendix}

\subsection{Brillouin scattering in one dimension}
Let consider the scattering of photons by phonons in one dimension.
Let consider the process of photon scattering on a phonon, i.e., the anti-Stokes processes. 
Let $\omega_{a,s,0}=ck_{a,s,0}$ be energies of anti-Stokes, Stokes, and pump photons, respectively, and $\omega_{\phi_{a,s}}=ck_{\phi_{a,s}}$ be energies of the phonons involved in the anti-Stokes and the Stokes processes, respectively.
We get
\begin{gather}\label{En_Mom_a}
    ck_0+vk_{\phi_a}=ck_a,\ \ \ \vec{k}_0+\vec{k}_{\phi_a}=\vec{k}_a,
    \\ \nonumber
    k_a^2=k^2_0+2\frac{v}{c}k_0 k_{\phi_a}+\frac{v^2}{c^2}k_\phi^2,
    \\ \nonumber
    k_a^2=k_0^2\pm2 k_0 k_{\phi_a} + k_{\phi_a}^2,
    \\ \nonumber
    \left(1-\frac{v^2}{c^2}\right)k_{\phi_a}^2
    \pm2\left(1\mp\frac{v}{c}\right)k_0 k_{\phi_a}=0.
\end{gather}

Then, $k_{\phi_a}=0$ or $k_{\phi_a}=\mp\frac{2k_0}{1\pm v/c}$.
The $k_{\phi_a}>0$ by definition.
Hence, $k_{\phi_a}=\frac{2k_0}{1- v/c}$, and $k_a=\frac{1+v/c}{1-v/c}k_0$.
Substituting these values into the second equation in Eq.~(\ref{En_Mom_a}), one can get that only one direction for $\vec{k}_{\phi_a}$ is allowed
\begin{gather}
    \vec{k}_{\phi_a}=-\frac{2k_0}{1-\frac{v}{c}} \frac{\vec{k}_0}{k_0},
    \ \ \ \ \ 
    \vec{k}_a=-\frac{1+\frac{v}{c}}{1-\frac{v}{c}} k_0 \frac{\vec{k}_0}{k_0},
\end{gather}
\begin{equation}
    \Delta\omega_a
    =ck_a-ck_0
    =\frac{2v/c}{1-v/c}\omega_0.
\end{equation}

It is seen that both the phonon that enters the anti-Stokes process and the exiting photon propagate against the direction of the initial photon.

Let consider the process of a photon's decay into a phonon and photon, i.e., the Stokes process.
\begin{gather}\label{En_Mom_s}
    ck_0=vk_{\phi_s}+ck_s,\ \ \ \vec{k}_0=\vec{k}_{\phi_s}+\vec{k}_s,
    \\ \nonumber
    k_s^2=k_0^2-2\frac{v}{c} k_0 k_{\phi_s} + \frac{v^2}{c^2} k_{\varphi_s}^2,
    \\ \nonumber
    k_s^2=k_0^2\pm2 k_0 k_{\phi_s} + k_{\phi_s}^2,
    \\ \nonumber
    \left(1-\frac{v^2}{c^2}\right)k_{\phi_s}^2
    \pm2\left(1\pm\frac{v}{c}\right)k_0 k_{\phi_s}=0.
\end{gather}

Then, $k_{\phi_s}=0$ or $k_{\phi_s}=\mp\frac{2k_0}{1\mp v/c}$.
The $k_{\phi_s}>0$ by definition.
Hence, $k_{\phi_s}=\frac{2k_0}{1+v/c}$, and $k_s=\frac{1-v/c}{1+v/c} k_0 $.
Substituting these values into the second equation of Eq.~(\ref{En_Mom_s}), one can get that only one direction for $\vec{k}_{\phi_s}$ is allowed
\begin{gather}
    \vec{k}_{\phi_s}=\frac{2k_0}{1+\frac{v}{c}} \frac{\vec{k}_0}{k_0},
    \ \ \ \ \ 
    \vec{k}_s=-\frac{1-\frac{v}{c}}{1+\frac{v}{c}} k_0 \frac{\vec{k}_0}{k_0},
\end{gather}
\begin{equation}
    \Delta\omega_s
    =ck_s-ck_0
    =-\frac{2v/c}{1+v/c}\omega_0.
\end{equation}

It is seen that the phonon that enters the Stokes process is co-directed with the initial photon, while the photon that exits the process is directed against them.

\subsection{Derivation of the used master equation}
Let consider the interaction of one ring resonator from the system in the main text with its phonon reservoir.
Let denote here $\hat{H}_S=\omega_s \hat{b}_s^\dag \hat{b}_s + \omega_a \hat{b}_a^\dag \hat{b}_a$, i.e., the Hamiltonian of the OQS, $\hat{H}_R=\sum_q \omega_q \hat{c}_q^\dag \hat{c}_q$, i.e., the Hamiltonian of a phonon reservoir, and the Hamiltonian of their interaction
\begin{gather}\label{HSR}
    \hat{H}_{SR}(t)
    =
    \sum_{q} \chi_q (\hat{b}_a^\dag \hat c_q e^{-i\omega_0 t} + \hat c_q^\dag \hat{b}_a e^{i\omega_0 t})
    \\ \nonumber
    +\sum_{q} \theta_q (\hat{b}_s^\dag \hat c_q^\dag e^{-i\omega_0 t} + \hat c_q \hat{b}_s  e^{i\omega_0 t})=
    \\ \nonumber
    =
    \hat{b}_a^\dag \hat{R}_1 (t) + \hat{b}_a \hat{R}_1^\dag (t)
    +\hat{b}_s^\dag \hat{R}_2 (t) + \hat{b}_s \hat{R}_2^\dag (t)
    =
    \sum_{v=1}^4 \hat{s}_v \hat{r}_v(t).
\end{gather}
Here $\hat{R}_1(t)=\sum_q\chi_q \hat{c}_q e^{-i \omega_0 t}$, $\hat{R}_2(t)=\sum_q\theta_q \hat{c}_q^\dag e^{-i \omega_0 t}$, terms in the last sum are just terms from the previous sum with $\hat{s}_v$ denoting all operators $\hat{b}$ and $\hat{b}^\dag$. 
Note that $\hat{H}=\hat{H}_S + \hat{H}_R + \hat{H}_{SR}(t)$.
In the interaction picture, $\dot{\hat{\tilde{\rho}}}(t)=i[\hat{\tilde{\rho}}(t),\hat{\tilde{H}}_{SR}(t)]$ ($\tilde{\cdot}$ denotes interaction picture)~\cite{shishkov2019relaxation}.
Considering $\chi_q$, $\theta_q$ to be small parameters of the order of $\varepsilon$, we get the expansion for the density matrix $\hat{\tilde{\rho}}(t)=\hat{\tilde{\rho}}_0(t)+\hat{\tilde{\rho}}_1(t)+\hat{\tilde{\rho}}_2(t) + O(\varepsilon^3)$, $\hat{\tilde{\rho}}_0(t)\sim 1$, $\hat{\tilde{\rho}}_1(t)\sim\varepsilon$, $\hat{\tilde{\rho}}_2(t)\sim\varepsilon^2$.
Then $\dot{\hat{\tilde{\rho}}}_0(t)=0$, $\hat{\tilde{\rho}}_0(t)=\hat{\tilde{\rho}}_0(0)\equiv \hat{\rho}_0$,
$\dot{\hat{\tilde{\rho}}}_1(t)=i[\hat{\rho}_0,\hat{\tilde{H}}_{SR}(t)]$,
\begin{equation}\label{p2}
    \hat{\tilde{\rho}}_{2}(t)
    =
    -\int\limits_0^t dt_1\int\limits_0^{t_1} dt_2
        [[\hat{\rho}_0,\hat{\tilde{H}}_{SR}(t_2)],\hat{\tilde{H}}_{SR}(t_1)].
\end{equation}
Let consider that the phonon reservoir is in thermal state $\hat{\rho}_{th}=e^{\hat{H}_R/T}/{\rm tr}_R(e^{\hat{H}_R/T})$ (Born approximation).
Hence, $\hat{\tilde{\rho}}(t)
=\hat{\rho}_S(t) \otimes\hat{\rho}_{th}
=(\hat{\tilde{\rho}}_{S0}(t)+\hat{\tilde{\rho}}_{S1}(t)+\hat{\tilde{\rho}}_{S2}(t) + O(\varepsilon^3))\otimes\hat{\rho}_{th}$, where $\hat{\rho}_S(t)$ is the reduced density matrix from the main text.
Applying trace operation over reservoir's degrees of freedom to both sides of Eq.~(\ref{p2}), we get $\dot{\hat{\tilde{\rho}}}_{S1}(t)=0$,
\begin{gather}\label{ps2}
    \hat{\tilde{\rho}}_{S2}(t)
    =
    -\sum_{v=1}^4\sum_{w=1}^4\int\limits_0^t dt_1\int\limits_0^{t_1} dt_2
        \\ \nonumber
        \Big(
            \hat{\rho}_0\hat{\tilde{s}}_v(t_2) \hat{\tilde{s}}_w(t_1) f_{vw}(t_2,t_1)
            -\hat{\tilde{s}}_v(t_2)\hat{\rho}_0\hat{\tilde{s}}_w(t_1) f_{wv}(t_1,t_2)
            \\ \nonumber
            -\hat{\tilde{s}}_w(t_1)\hat{\rho}_0 \hat{\tilde{s}}_v(t_2) f_{vw}(t_2,t_1)
            +\hat{\tilde{s}}_w(t_1)\hat{\tilde{s}}_v(t_2) \hat{\rho}_0 f_{wv}(t_1,t_2)  
        \Big).
\end{gather}
Here the function $f_{vw}(t_2,t_1)\equiv{\rm tr}_R(\hat{\rho}_{th}\hat{r}_v(t_2)\hat{r}_w(t_1))$,
$f_{11}=f_{22}=f_{33}=f_{44}=f_{14}=f_{41}=f_{23}=f_{32}=0$.
Indices $v,w$ enumerate terms in Eq.~(\ref{HSR}).
Then
\begin{gather}
    f_{12}(t_2,t_1)=\sum_q \chi_q^2 (n(\omega_q)+1)e^{-i(\omega_q+\omega_0)\tau}, \\ \nonumber
    f_{12}(t_1,t_2)=\sum_q \chi_q^2 (n(\omega_q)+1) e^{i(\omega_q+\omega_0)\tau}, \\ \nonumber
    f_{21}(t_2,t_1)=\sum_q \chi_q^2 n(\omega_q) e^{i(\omega_q+\omega_0)\tau}, \\ \nonumber
    f_{21}(t_1,t_2)=\sum_q \chi_q^2 n(\omega_q) e^{-i(\omega_q+\omega_0)\tau}, 
\end{gather}
\begin{gather}
    f_{13}(t_2,t_1)=\sum_q \chi_q \theta_q (n(\omega_q)+1)e^{-i\omega_q \tau}e^{-i\omega_0 \zeta}, \\ \nonumber
    f_{13}(t_1,t_2)=\sum_q \chi_q \theta_q (n(\omega_q)+1) e^{i\omega_q \tau}e^{-i\omega_0 \zeta},  \\ \nonumber
    f_{31}(t_2,t_1)=\sum_q \chi_q \theta_q n(\omega_q) e^{i\omega_q \tau}e^{-i\omega_0 \zeta},  \\ \nonumber
    f_{31}(t_1,t_2)=\sum_q \chi_q \theta_q n(\omega_q) e^{-i\omega_q \tau}e^{-i\omega_0 \zeta} , 
\end{gather}
\begin{gather}
    f_{42}(t_2,t_1)=\sum_q \chi_q \theta_q (n(\omega_q)+1)e^{-i\omega_q \tau}e^{i\omega_0 \zeta}, \\ \nonumber
    f_{42}(t_1,t_2)=\sum_q \chi_q \theta_q (n(\omega_q)+1) e^{i\omega_q \tau}e^{i\omega_0 \zeta},  \\ \nonumber
    f_{24}(t_2,t_1)=\sum_q \chi_q \theta_q n(\omega_q) e^{i\omega_q \tau}e^{i\omega_0 \zeta},  \\ \nonumber
    f_{24}(t_1,t_2)=\sum_q \chi_q \theta_q n(\omega_q) e^{-i\omega_q \tau}e^{i\omega_0 \zeta} , 
\end{gather}
\begin{gather}
    f_{43}(t_2,t_1)=\sum_q \theta_q^2 (n(\omega_q)+1)e^{-i(\omega_q-\omega_0)\tau}, \\ \nonumber
    f_{43}(t_1,t_2)=\sum_q \theta_q^2 (n(\omega_q)+1) e^{i(\omega_q-\omega_0)\tau}, \\ \nonumber
    f_{34}(t_2,t_1)=\sum_q \theta_q^2 n(\omega_q) e^{i(\omega_q-\omega_0)\tau}, \\ \nonumber
    f_{34}(t_1,t_2)=\sum_q \theta_q^2 n(\omega_q) e^{-i(\omega_q-\omega_0)\tau}. 
\end{gather}
Here $\tau=t_2-t_1$, $\zeta=t_2+t_1$.
Also

\begin{gather}
\nonumber
    \hat{\tilde{s}}_1 (t_2) \hat{\tilde{s}}_2 (t_1)
    =
    \hat{b}_a^\dag \hat{b}_a e^{i\omega_a \tau}, 
    \ \ \ 
    \hat{\tilde{s}}_1 (t_1) \hat{\tilde{s}}_2 (t_2)
    =
    \hat{b}_a^\dag \hat{b}_a e^{-i\omega_a \tau},
    \\ 
    \hat{\tilde{s}}_1 (t_2) \hat{\tilde{s}}_3 (t_1)
    \approx
    \hat{b}_a^\dag \hat{b}_s^\dag e^{i\omega_0 \zeta + i\omega_\varphi \tau}, 
    \\ \nonumber
    \hat{\tilde{s}}_2 (t_1) \hat{\tilde{s}}_4 (t_2)
    \approx
    \hat{b}_a \hat{b}_s e^{-i\omega_0 \zeta +i\omega_\varphi \tau},
    \\ \nonumber
    \hat{\tilde{s}}_3 (t_2) \hat{\tilde{s}}_4 (t_1)
    =
    \hat{b}_s^\dag \hat{b}_s e^{i\omega_s \tau}, 
    \ \ \ 
    \hat{\tilde{s}}_3 (t_1) \hat{\tilde{s}}_4 (t_2)
    =
    \hat{b}_s^\dag \hat{b}_s e^{-i\omega_s \tau}.    
\end{gather}
The coupling constants $\chi_q\sim\int dl e^{-i k_a l} e^{-i k_q l} e^{-i k_0 l}$, $\theta_q\sim\int dl e^{-i k_s l} e^{-i k_q l} e^{-i k_0 l}$~\cite{haus2002coupled,yeh2008essence,haus2003coupled}.
Hence, $\chi_q \theta_q=0$ (see previous part of Appendix). 
This considerably reduces the number of terms.

We can use integration instead of the sum over $q$ in equations for $f_{vw}$, i.e., $\sum_q\ldots=\int_0^{+\infty}d\omega g(\omega)\ldots$.
In Markov approximation~\cite{breuer2002theory,carmichael2009open}
\begin{equation}
    \int_0^{t_1}dt_2=\int_{-t_1}^0 d\tau\overset{\text{Markov approx.}}{=}\int_{-\infty}^0 d\tau.
\end{equation}

After integration, terms with $f_{12}$ and $f_{21}$ deliver anti-Stokes secular terms of the master equation from the main text, and terms with $f_{43}$ and $f_{34}$ deliver Stokes secular terms of the master equation
\begin{gather}
    \frac{\partial \hat \rho}{\partial t}
    =
    i\left[
        \hat \rho,
        \omega_a \hat{b}_a^\dag \hat{b}_a
        +\omega_s \hat{b}_s^\dag \hat{b}_s
    \right]
    \\ \nonumber
    +\gamma(\omega_\phi)
    \left(
        \frac{n(\omega_\phi,T)+1}{2}\hat L[\hat b_a,\hat b^\dag_a] 
        +\frac{n(\omega_\phi,T)}{2}\hat L[\hat b^\dag_a,\hat b_a]
    \right)
    \\ \nonumber
    +\gamma(\omega_\phi)
    \left(
        \frac{n(\omega_\phi,T)+1}{2}\hat L[\hat b_s^\dag,\hat b_s] 
        +\frac{n(\omega_\phi,T)}{2}\hat L[\hat b_s,\hat b^\dag_s]
    \right).
\end{gather}
Here $\omega_\phi=\omega_a-\omega_0\approx\omega_s-\omega_0$ is the frequency of phonons involved in the Brillouin scattering, superoperator $\hat L [\hat X, \hat Y] = 2\hat X \hat \rho_S \hat Y-\hat Y \hat X \hat\rho_S -\hat\rho_S \hat Y \hat X$, $n(\tilde\omega,T)=(e^{\tilde\omega/T}-1)^{-1}$ is the occupancy of the $k$-th reservoir at frequency $\tilde\omega$ ($T$ is the temperature of the reservoir), $\gamma(\tilde\omega) = \pi g_k\left(\tilde\omega\right) |\chi_{\mathbf{q}}|^2$ is the Brillouin scattering rate, where $g\left(\tilde\omega\right)$ is the density of states in the phonon reservoir, and $\chi_{\mathbf{q}}$ is the coupling constant $\chi_{q}$ at frequency $\omega_{q}=\tilde\omega$.

\subsection{Stationary solution of the Eqs.~(\ref{Diff_occ}) from the main text}

Let use here the notation $\langle\hat{a}_a^\dag \hat{a}_a\rangle=A$, $\langle\hat{b}_{k,a}^\dag \hat{b}_{k,a}\rangle=B_k$, $\langle\hat{a}_a^\dag \hat{b}_{k,a}\rangle=x_k+iy_k$, $\langle\hat{b}_{j,a}^\dag \hat{b}_{k,a}\rangle=\alpha_{jk}+i\beta_{jk}$, $\gamma_a\equiv\gamma_a(\omega)$, $\gamma_k (\omega_\phi)\equiv\gamma_k$, $n_k\equiv n_k(\omega_\phi,T_k)$.
Then (further, we drop out the index $a$ as we consider the dynamics of anti-Stokes modes only)
\begin{gather}
    \dot{A}=2\sum_{k=1}^N \Omega_k y_k - \gamma_a A, \\ \nonumber
    \dot{B}_k=-2 \Omega_k y_k - I\gamma_k B_k +I\gamma_k n_k, \\ \nonumber
    \dot{x}_k
    =
    - \sum_{j\ne k}^N \Omega_j \beta_{jk} -\frac{\gamma_a + I\gamma_k}{2} x_k, \\ \nonumber
    \dot{y}_k
    =
    \Omega_k (B_k-A) + \sum_{j\ne k}^N \Omega_j \alpha_{jk} 
    -\frac{\gamma_a + I\gamma_k}{2} y_k,\\ \nonumber
    \dot{\alpha}_{jk}=-\Omega_j y_k - \Omega_k y_j - I\frac{\gamma_j + \gamma_k}{2} \alpha_{jk},\\ \nonumber
    \dot{\beta}_{jk}=\Omega_j x_k - \Omega_k x_j - I\frac{\gamma_j + \gamma_k}{2} \beta_{jk}.
\end{gather}

At stationary state, all time derivatives from above equations equal zero.
Hence,
\begin{gather}
    \overline{\alpha}_{jk}
    =
    \frac{-2/I}{\gamma_j+\gamma_k}(\Omega_j \overline{y}_k + \Omega_k \overline{y}_j), \\ \nonumber
    \overline{\beta}_{jk}
    =
    \frac{2/I}{\gamma_j+\gamma_k}(\Omega_j \overline{x}_k - \Omega_k \overline{x}_j), \\ \nonumber
    \overline{y}_k 
    \left(\frac{\gamma_a+I\gamma_k}{2}+\sum_{j\ne k}^N \frac{2\Omega_j^2/I}{\gamma_j+\gamma_k}\right)
    +\sum_{j\ne k}^N\frac{2\Omega_j \Omega_k/I}{\gamma_j+\gamma_k} \overline{y}_j=\\ \nonumber
    =
    \Omega_k (\overline{B}_k - \overline{A}), \\ \nonumber
    \overline{x}_k\left(\frac{\gamma_a+I\gamma_k}{2}+\sum_{j\ne k}^N \frac{2\Omega_j^2/I}{\gamma_j+\gamma_k}\right)
    -\sum_{j\ne k}^N\frac{2\Omega_j \Omega_k/I}{\gamma_j+\gamma_k} \overline{x}_j
    =
    0,
    \\ \nonumber
    \overline{B}_k = -\frac{2\Omega_k}{I\gamma_k}\overline{y}_k + n_k.
\end{gather}
Here $\overline{\,\cdot\,}$ denotes the stationary value of the variable.
It is natural to assume that all $\gamma_k$ are of the same order, i.e., $\gamma_k\sim\gamma_\phi$.
Let introduce $\overline{u}_k$ as $\overline{y}_k\equiv2\Omega_k \overline{u}_k/(\gamma_a+I\gamma_\phi)$, then
\begin{gather}
    \Omega_k \overline{u}_k 
    + \frac{2\Omega_k/I\gamma_\phi}{\gamma_a+I\gamma_\phi}
    \sum_{j\ne k}^N\Omega_j^2(\overline{u}_k+\overline{u}_j)
    =
    \Omega_k (\overline{B}_k - \overline{A}).
\end{gather}
If $\varepsilon\sim\Omega_k/I\gamma_\phi \ll1$ and $\gamma_a\ll \gamma_\phi$, then the second term in this equation is of the order $\Omega_k^2/I\gamma_\phi \max(I\gamma_\phi,\gamma_a)\sim\varepsilon^2$.
We can neglect this term when $N\varepsilon^2\ll 1$.
Hence,
\begin{gather}
    \overline{y}_k\approx\frac{2\Omega_k}{\gamma_a+I\gamma_k}(\overline{B}_k - \overline{A}), \\ \nonumber
    \overline{B}_k\approx n_k + \frac{4\Omega_k^2 \overline{A}}{I\gamma_k (\gamma_a+I\gamma_k)}, \\ \nonumber
    \overline{A}\approx\frac{\sum_{k=1}^N r_k \overline{B}_k}{\sum_{k=1}^N r_k+\gamma_a}\approx \frac{\sum_{k=1}^N r_k n_k}{\sum_{k=1}^N r_k + \gamma_a}.
\end{gather}
Here, $r_k=4\Omega_k^2/(\gamma_a+I\gamma_\phi)$.
As we have neglected $\varepsilon^2$ terms, we should check that the achieved solutions include terms only of lower orders.
The $\overline{B}_k\sim n_k+\varepsilon^2 \overline{A}$, then
$\overline{y}_k\sim \varepsilon (\overline{B}_k - \overline{A})$ and
\begin{equation}
    \overline{A}\sim 
    \frac{N\varepsilon^2 n_k}{N\varepsilon^2 +{\gamma_a}/{I\gamma_\phi}}.
\end{equation}
Thus, the $\gamma_a\ll I\gamma_\phi$ should be fulfilled. Otherwise $\overline{A}$ would be a term of $\varepsilon^2$ order. 
If this condition is fulfilled, the influence of phonon reservoirs establishes stationary occupancies of the ring resonators' optical modes $\overline{\langle\hat{b}_k^\dag \hat b_k\rangle}\approx n_k(\omega_\phi)$.
Simultaneously with that, the stationary occupancy of the anti-Stokes mode $\overline{I}_a=\overline{\langle\hat{a}_a^\dag \hat{a}_a\rangle}$ is established:
\begin{equation}
     \overline{I}_a=\overline{\langle\hat{a}_a^\dag\hat{a}_a\rangle}
     \approx
     \frac{\sum\limits_{k=1}^N r_k (\omega_\phi) n_k(\omega_\phi)}{\sum\limits_{k=1}^N r_k(\omega_\phi) +\gamma_a (\omega)}.
\end{equation}
Here $r_k(\omega_\phi)=4\Omega_k^2/ (I\gamma_k(\omega_\phi)+\gamma_a (\omega))$.


\end{document}